\documentclass[journal=acs photonics,manuscript=article]{achemso}

\usepackage{amsmath}
\usepackage{geometry}%
\usepackage{amsfonts}%
\usepackage{amssymb}%
\usepackage{graphicx}

\usepackage{xcolor}
\author{S. Ali Hassani Gangaraj}
\affiliation[Unknown University]
{Optical Physics Division, Corning Research and Development, Sullivan Park, Corning, New York, 14831, USA}
\alsoaffiliation[Second University]
{Department of Electrical and Computer Engineering, University of Wisconsin-Madison, Madison, WI 53706, USA}
\email{ali.gangaraj@gmail.com}

\author{Boyuan Jin}
\affiliation[Unknown University]
{Department of Electrical and Computer Engineering, University of Nebraska-Lincoln, Lincoln, NE 68588, USA}
\email{byjin328@huskers.unl.edu}

\author{Christos Argyropoulos}
\affiliation[Unknown University]
{Department of Electrical Engineering, The Pennsylvania State University, University Park, PA, 16802, USA}
\alsoaffiliation[Second University]
{Department of Electrical and Computer Engineering, University of Nebraska-Lincoln, Lincoln, NE 68588, USA}
\email{cfa5361@psu.edu}

\author{Francesco Monticone}
\affiliation[Unknown University]
{School of Electrical and Computer Engineering, Cornell University, Ithaca, New York 14853, USA}
\email{francesco.monticone@cornell.edu}

\title[An \textsf{achemso} demo]
{Enhanced nonlinear optical effects in drift-biased nonreciprocal graphene plasmonics}

\begin{document}

%\begin{tocentry}
%\includegraphics[ width=1\linewidth, keepaspectratio]{TOC.png}

%\end{tocentry}

\begin{abstract}

Nonlinear light-matter interactions are typically enhanced by increasing the local field and its interaction time with matter. Conventional methods to achieve these goals are based on resonances or slow-light effects. However, these methods suffer from various issues, including narrow operational bandwidths, large footprints, and material absorption. An interesting alternative approach to enhance the local field is offered by nonreciprocal systems: by blocking the path of a unidirectional wave in a terminated nonreciprocal waveguiding structure, broadband electromagnetic fields can be drastically enhanced and localized near the termination. This approach was previously studied only in three-dimensional gyrotropic material platforms where the need for external magnets and bulky materials make it less practical. Here, instead, we employ a magnet-free mechanism to break reciprocity in 2D plasmonic materials, e.g., graphene. Specifically, we employ high-speed drifting electrons on a voltage-biased graphene sheet to lift the forward/backward degeneracy of the surface plasmon-polariton dispersion, creating modes with different propagation properties parallel and antiparallel to the current. We show that controllable, asymmetric, and intense field hot-spots are generated at the edges of a suitably terminated graphene metasurface. We then theoretically demonstrate that such asymmetric field hot-spots offer an effective solution to enhance third-order nonlinear optical effects. As an example, we predict that, using realistic values of drift velocity, high third-harmonic conversion efficiencies of up to 0.3 percent are achievable around the plasmon resonance frequencies.

\end{abstract}


\section{Introduction}

Nonlinear light-matter interactions are typically weak, but they can be boosted by enhancing the local electromagnetic field and and by increasing the interaction time between photons and matter \cite{Khurgin} through various strategies. Among different platforms, plasmonic materials have been extensively studied as they enable strong forms of light confinement and enhancement thanks to the coupling of electromagnetic field and free-charge-density oscillations \cite{Schuller}. In plasmonics, traditional methods to achieve strong local field enhancement are: (i) localized surface plasmon resonances \cite{Kauranen,Alu_1,Alu_2}, or (ii) slow-light effects in adiabatically tapered structures, where the energy carried by the surface plasmon-polaritons (SPPs) accumulates and forms a ``hot-spot'' at the tip of an elongated plasmonic taper \cite{Stockman,Davoyan,Alu_3}. However, these methods suffer from either narrow operational bandwidths due to resonant effects or the need for long adiabatically tapered micro-nano-structures. Recently, a different strategy has been proposed to achieve broadband field localization and enhancement, which can be used to boost nonlinear effects and efficiently generate second and third harmonics. This method is based on exploiting the extreme response of a terminated nonreciprocal plasmonic platform that supports inherently unidirectional surface modes, i.e., SPPs that are allowed to propagate along a certain direction but not in the opposite one, within a relatively broadband unidirectional frequency window. Since the SPP cannot reflect back, its energy accumulates at the termination and forms an intense and broadband electric field hot-spot, without the need for a long adiabatic taper or a localized resonance \cite{Hassani_PRApplied, Mann}. This strategy breaks the above-mentioned conventional trade-offs between field enhancement, bandwidth, and size. However, in previous studies, reciprocity was broken via bulky gyrotropic materials, which requires a strong external magnetic bias \cite{Hassani_diffractionless,Hassani_Optica,Hassani_EP,Hassani_AWPL}. Hence, their large footprint, weak gyrotropic response at optical frequencies, and challenges to their integration in photonic and electromagnetic systems make them less appealing. 

In this work, we employ a magnet-free mechanism to break reciprocity and achieve field localization and enhanced light-matter interactions in 2D materials supporting asymmetric SPP propagation. We also investigate how this approach can provide exciting opportunities for enhanced nonlinear interactions. Among a large range of plasmonic materials available \cite{Tassin}, graphene in particular has received considerable attention as a promising 2D platform in both classical electrodynamics and quantum optics, with a diverse range of applications from large Purcell enhancement of emission to transformation optics \cite{Low_1,Abajo_1,Hanson,Abajo_2,Hanson_2,Hassani_PRA,Sounas_graphene,Vakil,Hassani_TAP}. Recent experiments have demonstrated that graphene also exhibits strong third-order optical nonlinearity at terahertz and infrared frequencies \cite{Kumar,Hong,Guo}. More specifically, the third-order nonlinear susceptibility of graphene has been measured to be several orders of magnitude higher than the one of conventional bulk nonlinear dielectric materials, as well as noble metals \cite{Kumar,Zhang}. Moreover, since the third-order nonlinear polarization is proportional to the cubic power of the local electric field \cite{Boyd}, the overall nonlinear response can be boosted through an electric field enhancement, which is often achieved via plasmonic resonances \cite{Kauranen,Chris_1,Chris_2,Chris_3,Chirs_4,Chirs_5,Chirs_6,Chris_7, Chirs_8}. In this regard, graphene also offers a promising plasmonic platform that supports highly-confined, relatively long-range, and electrically tunable SPPs \cite{Jablan,Woessner,Basov_1,Basov_2}.

In addition to the possibility of tuning the carrier concentration in graphene through chemical doping and electrical gating, it is also possible to induce a drifting current on the graphene surface, which may enable additional degrees of freedom to control graphene plasmons. The effect of drifting electrons on propagating SPPs can be qualitatively understood in a simple way: SPPs are collective charge-density oscillations coupled to photons and, therefore, they are dragged or opposed by drifting electrons, which causes SPPs to see different optical properties when propagating parallel or antiparallel to the drift velocity. More precisely, graphene conductivity becomes nonlocal (i.e., direction-dependent or spatially dispersive) in the presence of drifting electrons, with this nonlocal response originating from the Doppler frequency shift produced by the movement of electrons \cite{Landau-1}, $ \omega \rightarrow \omega - \boldsymbol{k} \cdot \boldsymbol{v}_d $, where $\boldsymbol{v}_d$ is the drift velocity. As a result, the eigenmodal solution depends on the sign of the wavevector $ \boldsymbol{k} $, which is a clear sign of nonreciprocity. Without this nonlocal effect, the forward- and backward-propagating SPPs have the same properties at any given frequency. This magnet-free method of breaking reciprocity recently received significant attention in 3D or 2D plasmonic waveguiding systems \cite{Bliokh,Hassani_ACS,Mario-ACS,Mario-Active,Mario-Negative-Landau,Mario-Nonlocal,Mario-Plasmonics,Collimated-SPP,Stauber,Levitov,Polini,Wenger}. To achieve a strong nonreciprocal response, large drift velocities are required, which is impractical in 3D solid-state materials, e.g., metals and semiconductors. Instead, graphene supports ultra-high current densities \cite{Mishchenko} and its carrier drift velocity can be comparable with the Fermi velocity, $v_F \approx 10^6 ~m/s$ \cite{Shishir,Dorgan,Ozdemir,Ramamoorthy,Yamoah}, which is orders of magnitude larger than the achievable drift velocities in metals such as gold \cite{Bliokh} or even high-mobility semiconductors like indium antimonide \cite{InSb}. Thus, graphene can be considered a uniquely well-suited 2D plasmonic platform for these purposes and, indeed, recent experiments have shown that strong drift-induced nonreciprocity is achievable at optical frequencies \cite{Basov_Fizeau,Wang_Fizeau}.

Here, we discuss asymmetric SPP propagation on graphene in the presence of drifting electrons and investigate their behavior in a terminated graphene waveguiding structure. Interestingly, the nonreciprocal propagation properties imply the emergence of intense asymmetric field hot-spots at the terminations, where the electromagnetic fields are extremely localized and enhanced. By leveraging these properties, we theoretically show that such asymmetric hot-spots offer an intriguing solution to strengthen the nonlinear optical response of a graphene structure, which may lead to a large enhancement of, for example, third-harmonic generation effects.

\section{Asymmetric Plasmon Propagation on Drift-Biased Graphene}

\begin{figure}[h!]
	\begin{center}
		\noindent \includegraphics[width=6.3in]{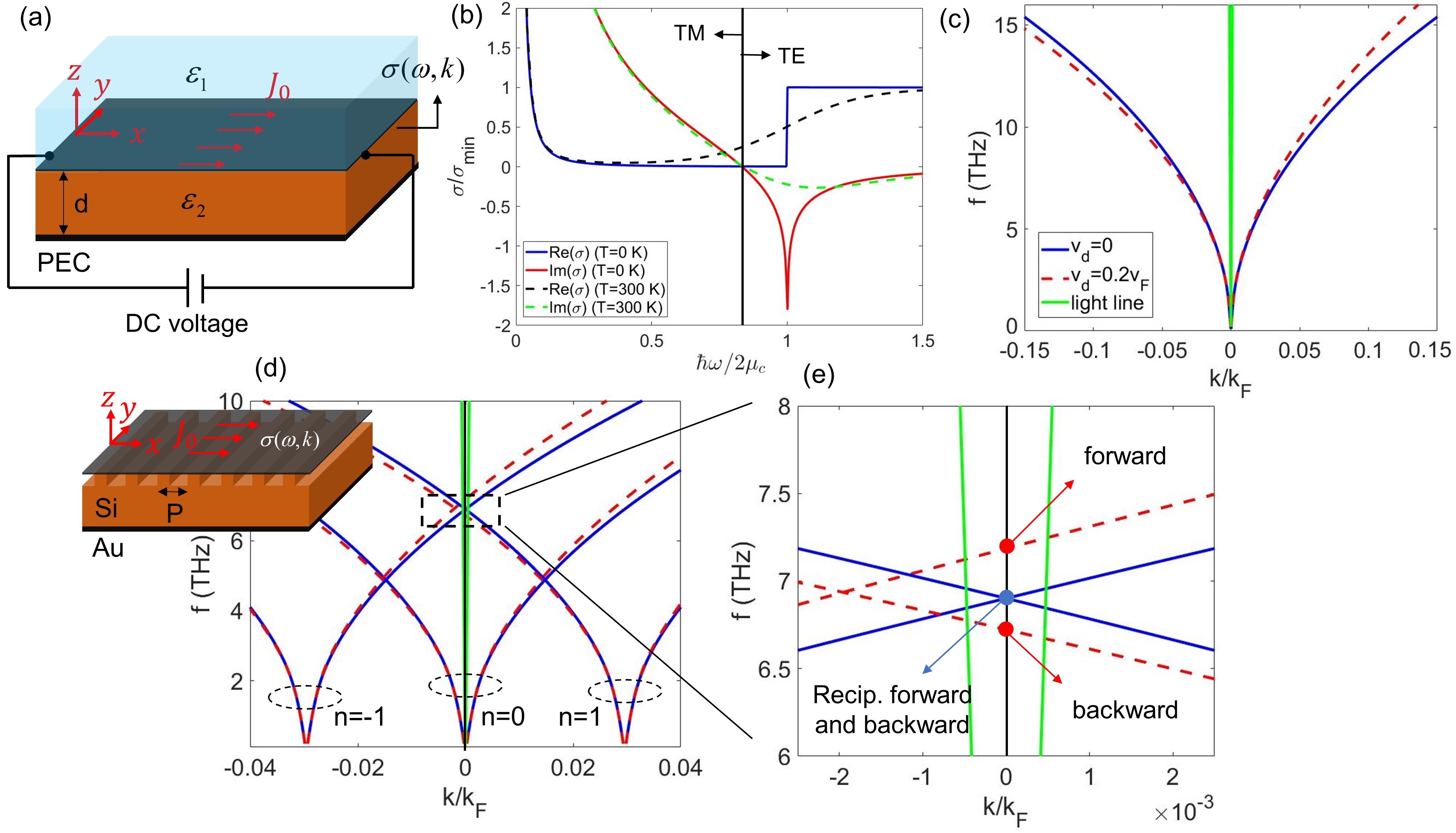}
	\end{center}
	\caption{(a) A graphene sheet supporting drifting electrons. A longitudinal voltage induces an electron drift in a graphene layer placed between a semi-infinite dielectric with permittivity $ \epsilon_1 $ and a grounded dielectric with $ \epsilon_2 $ (PEC indicates a perfect electric conductor). (b) Graphene's conductivity at room temperature $T=300$ K and $ T= 0 $ K. The conductivity has been rescaled by $\sigma_{min} = \pi e^2/2h$. The chemical potential is $ \mu_c = 0.2 $ eV, and $ 1 / \Gamma = 0.33 $ ps, $ 1/\Gamma' = 0.0658$ ps. These values are assumed throughout our work unless otherwise noted. (c) Dispersion diagram of the SPPs supported by the graphene layer for different drift velocities. Graphene is placed between vacuum and a grounded silicon substrate with thickness $ d = 2 ~\mu$m. (d) Dispersion diagram of SPPs supported by a graphene sheet over a grating silicon substrate with grating periodicity $ P = 0.7 ~\mu$m. This panel shows the fundamental, $ n = 0$, and first, $ n = \pm 1 $, space harmonics. As can be seen, the first harmonics enter the light cone, implying that they can be excited by an incident propagating plane wave. (e) Zoomed-in view of panel (d) around the light line (solid green line).   }\label{fig1}
\end{figure}

We start by considering a laterally infinite graphene sheet lying on the $x-y$ plane and sandwiched between a semi-infinite dielectric with permittivity $ \epsilon_1 $ and a grounded dielectric with finite thickness $d$ and permittivity $ \epsilon_2 $, as shown in Fig. \ref{fig1}(a). Graphene can be modeled as an infinitesimally thin 2D surface characterized by a frequency-dispersive surface conductivity $ \sigma(\omega) $ \cite{Gusynin, Hanson}:
\begin{equation}\label{s_gr}
	\sigma(\omega) = \frac{ ie^2 k_B T }{ \pi \hbar^2(\omega + i\Gamma) } \left[ \frac{\mu_c}{k_B T} + 2 \mathrm{ln} \left( e^{-\frac{\mu_c}{k_B T}} + 1 \right) \right] + \frac{i e^2 ( \omega+ i \Gamma' )}{\pi \hbar^2} \int_{0}^{\infty} \frac{ f_d(-\mathcal{E}) - f_d(\mathcal{E})  }{ (\omega + i\Gamma')^2 - 4(\mathcal{E}/\hbar)^2  } d\mathcal{E},
\end{equation}
where $ \omega $ is the frequency, $ \mu_c $ is the chemical potential
(Fermi level), $ \Gamma $ and $ \Gamma' $ are the intraband and interband scattering rates, respectively, $ T$ is the temperature, $e$
is the electron charge, $k_B$ is Boltzmann’s constant, and $ f_d(\mathcal{E}) = (e^{(\mathcal{E}-\mu_c)/k_BT} + 1)^{-1} $ is the Fermi-Dirac distribution. The first and second terms in Eq. (\ref{s_gr}) represent intraband and interband contributions, respectively. Equation (\ref{s_gr}) is valid for any temperature $T$. Figure \ref{fig1}(b) shows the graphene conductivity at room temperature ($ T =300 $ K) and $ T = 0  $ K for $ \mu_c = 0.2 $ eV and $ 1 / \Gamma = 0.33 $ ps and $ 1/\Gamma' = 0.0658$ ps. 

At low frequencies $ \hbar \omega \ll 2 \mu_c $, the intraband conductivity is dominant (Drude response), $ \mathrm{Im}(\sigma) > 0 $ (inductive surface reactance), and transverse-magnetic (TM) SPPs are supported by graphene. As the frequency increases, the Drude part of the conductivity falls off, and interband absorption becomes important. A loosely confined transverse-electric (TE) surface plasmon wave can propagate at higher frequencies where $ \mathrm{Im}(\sigma) < 0 $ (capacitive surface reactance). Therefore, as indicated in Fig. \ref{fig1}(b), the supported surface mode is TM to the left of the discontinuity and TE to the right . Our study concerns TM SPPs at $ \hbar \omega \ll 2 \mu_c $, where, as is clear from Fig. \ref{fig1}(b), low and high temperature models essentially coincide. At low enough temperatures, $ T \rightarrow 0 $, the graphene conductivity becomes virtually temperature-independent and takes a simpler form, which is used in our study to simplify the calculations.

If a DC voltage is longitudinally applied on the graphene layer, a current of drifting electrons is generated with drift velocity $ \boldsymbol{v}_d = \boldsymbol{J_0}/ne $, where $n$ is the electron density and $\boldsymbol{J_0}$ is the current density. As explained in the Introduction section, the presence of drifting electrons produces a Doppler frequency shift, leading to a frequency dispersive and spatially dispersive conductivity, $ \sigma(\omega, k) $, which can be written as \cite{Mario-Negative-Landau,Mario-Nonlocal,Mario-ACS},
\begin{equation}
	\sigma(\omega, k) = \frac{\omega}{ \omega - \boldsymbol{k} \cdot \boldsymbol{v}_d } \sigma_0( \omega - \boldsymbol{k} \cdot \boldsymbol{v}_d ),
\end{equation}
where $ \boldsymbol{k} $ denotes the SPPs wave-vector and $ \sigma
_0 $ is graphene’s conductivity in the absence of drifting electrons.

To study the effect of the drift-current bias, we analyze the supported TM SPPs. Assuming for the moment SPP propagation in the $x$-direction, a TM mode has nonzero components ($E_x,~ E_z,~H_y$). By solving Maxwell’s equations and applying the appropriate boundary conditions for graphene, $ \hat{z} \times (\boldsymbol{H}^+ - \boldsymbol{H}^-) = \sigma \boldsymbol{E}  $ and $ \hat{z} \times (\boldsymbol{E}^+ - \boldsymbol{E}^-) = 0  $, it can be shown that the dispersion equation of the TM SPPs supported by the structure shown in Fig. \ref{fig1}(a) can be written as:
\begin{equation}
	\frac{\gamma_1}{\epsilon_1} \left(  1 + e^{-2\gamma_2 d}   \right) - \frac{\gamma_2}{\epsilon_2} \left( 1 + \frac{ i \sigma(\omega, k) \gamma_1 }{\omega \epsilon_0 \epsilon_1}  \right) \left(  e^{-2\gamma_2 d} -1   \right) = 0
\end{equation}
where $ \gamma_i = \sqrt{k_x^2 - \epsilon_i (\omega/c)^2}, ~ i=1,~2 $. Evidently, without the biasing current, and assuming $ \epsilon_1 = \epsilon_2 = \epsilon_r $ and $ d \rightarrow \infty $, the above equation reduces to the well-known dispersion relation for graphene's TM SPPs \cite{Hanson_TM}, $ k_x = k_0 \sqrt{\epsilon_r  \left[  1 - \left( 2/\sigma \eta\right)^2  \right]   } $, where $ k_0 = \omega /c $ and $ \eta = \sqrt{\mu_0 / \epsilon_0 \epsilon_r} $. Figure \ref{fig1}(c) shows the dispersion relation of TM SPPs for different electron drift velocities. Here, it is assumed that graphene has been placed at the interface between silicon and vacuum. The silicon layer's thickness is $ d = 2 \mu $m. Graphene's parameters are written in the caption of Fig. \ref{fig1}. In the absence of drift current, the dispersion curve is formed by two symmetric branches (solid blue lines) indicating reciprocity and two identical counter-propagating modes. In contrast, if the graphene sheet is biased by drifting electrons the symmetry breaks (dashed red line), which indicates nonreciprocity and different propagation properties along the $+x$ and $-x$ directions. Here, it is assumed $ v_d = v_F/5 $, where $ v_F = c/300 $ is the Fermi velocity. Recent experiments \cite{Basov_Fizeau,Wang_Fizeau} have demonstrated that this level of drift velocity is achievable in graphene. We also would like to note that more sophisticated models that also account for the intrinsic nonlocality of graphene, the impact of heating, and other effects, as those in, e.g., Ref. \cite{Basov_Fizeau,Current_controlled}, may lead to more accurate predictions, without however qualitatively changing the effects predicted and discussed here. For example, in Ref. \cite{Basov_Fizeau}, it was found that significant Joule heating was not observed under similar conditions (similar drift velocity), and heating effects were observed to only produce a small discrepancy between the measurements and the theoretical predictions. Moreover, in Ref. \cite{Current_controlled} a graphene conductivity model including both intrinsic and drift-induced nonlocalities led to similar predictions for the dispersion diagram of TM SPPs supported by a current-biased graphene sheet, especially for the relatively small values of wavenumber relevant to our work. Further details are provided in the Supporting Information.

We note that the TM SPP dispersion curves lie out of the light cone (solid green lines in Fig. \ref{fig1}(c)) and, therefore, these modes cannot be excited by a plane wave propagating in free space. A practical configuration that allows the surface mode to be excited by an incident propagating wave can be realized using a grating coupler created by periodically corrugating the surface of the substrate, as in the inset of Fig. \ref{fig1}(d). In this case, the period $P$ is chosen such that the grating compensates the wavevector mismatch between the surface wave and the incident plane wave, i.e., $ k_{spp} = k_0 \cos (\theta) + 2\pi n / P $ where $ k_0 $ is the free-space wavenumber and $ \theta $ is the angle of incidence measured from the interface \cite{grating}. If the graphene sheet is placed at the interface between the silicon grating and vacuum, then it is possible to launch graphene SPPs using a propagating plane-wave excitation. Figure \ref{fig1}(d) shows the fundamental, $n = 0$ and first, $ n = \pm 1$, space harmonics. Here, it is assumed that the grating period of the silicon substrate is $ P = 0.7 ~ \mu$m and the grating region is laterally infinite. As seen in the figure, the first harmonics enter the light-cone and cross the vertical axis at $ k = 0 $, which means that SPPs may be excited by a plane wave illuminating the structure along the normal direction to the interface. A zoomed-in view around the light-cone is shown in Fig. \ref{fig1}(e) for different drift velocities (for simplicity of illustration, in these dispersion diagrams we neglect the coupling between different harmonics, which would lead to the opening of bandgaps at the crossing points; the simulations in the next sections are instead exact). In the absence of drifting electrons, the backward- and forward-propagating modes (their space harmonics) cross the $k=0$ axis at the same point (blue dot), which means that, if the structure is normally illuminated, we expect an absorption resonance around the frequency of the crossing point. In contrast, in the presence of drifting electrons, backward- and forward-propagating modes cross the $k=0$ axis at different frequencies (red dots), which implies that once the system is biased, the resonance splits into two distinct resonances, associated with nonreciprocal plasmons propagating parallel and antiparallel to the drifting electrons. It should be noted that higher-order harmonics, $ n  = 2, ~3, ~ 4, ~ \dots $, cross $ k = 0 $ at higher frequencies and are not shown here. In the next section, we show that this effect can lead to an asymmetric field enhancement at the edges of a finite-length terminated graphene structure, which may then be used to boost nonlinear light-matter interactions by orders of magnitude.

\section{Asymmetric Field Hot-Spots in a Terminated Drift-Biased Graphene Structure}

\begin{figure}[h!]
	\begin{center}
		\noindent \includegraphics[width=6.3in]{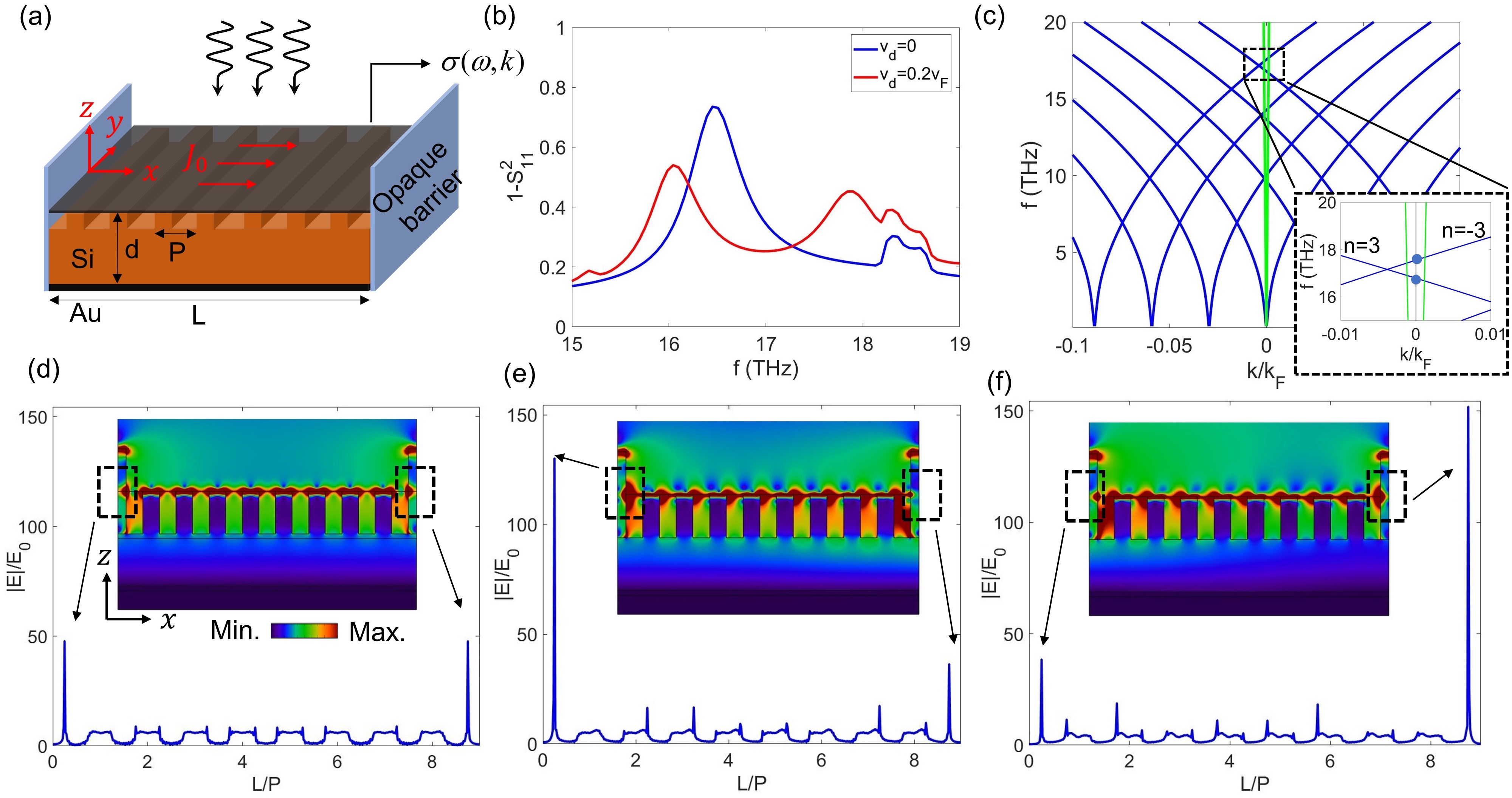}
	\end{center}
	\caption{ (a) A schematic view of the considered system. The graphene sheet is placed on a periodically corrugated silicon substrate with thickness $ d = 2 ~ \mu m$, with grating periodicity $ P = 0.7~ \mu m $ and depth $ h = 0.9 ~ \mu m $. The structure is supposed to be infinite along the $ y $-axis and finite along the $ x $-axis with $ L = 8.5 P $. A drifting current is assumed to be supported by graphene along the $x$-axis. The structure is terminated at the two ends by opaque barriers with permittivity following a Drude model with $ \epsilon_b = -2 - i0.1 $ at the frequency of the absorption peak in the non-biased case in panel (b). (b) Absorption as a function of incident frequency for two values of $v_d$. As $ v_d $ is increased the SPP resonance peak separates into two distinct resonances. (c) Dispersion diagrams of SPPs supported by a graphene sheet over a silicon substrate grating that is laterally infinite. This panel shows the first three space harmonics. The inset shows a zoomed-in view around the light lines (solid green lines). (d,e,f) Distribution of the electric field magnitude along the graphene sheet at the absorption peak frequency for (d) nonbiased case, (e) current-biased, lower frequency peak and (f) current-biased, higher frequency peak. The insets show spatial distributions of the electric field magnitude on the $xz$-plane for each case separately. }\label{fig2}
\end{figure}

In this section we investigate the potential of asymmetric SPPs supported by drift-biased graphene to create intense field hot-spots in a terminated graphene structure. The setup under consideration is shown in Fig. \ref{fig2}(a). A graphene sheet is placed on top of a corrugated silicon substrate. The grating periodicity and depth are $ P = 0.7 ~ \mu $m and $ h = 0.9 ~ \mu$m, respectively. The graphene sheet is considered to be infinite along the $y$-axis and finite ($L = 8.5P$) along the $x$-axis, with material parameters given in the caption of Fig. \ref{fig1}. The silicon thickness is $ d = 2~ \mu $m and is backed by a thin layer of gold. Graphene is terminated at the two ends along the $x$-axis by a lossy opaque material (e.g., a conducting material) with permittivity following a Drude model with $ \epsilon_b = -2 - i0.1 $ at the frequency of the resonance peak in the non-biased case, shown in Fig. \ref{fig2}(b). These barriers in Fig. \ref{fig2}(a) serve two purposes: enhancing the field hot-spot intensity at the terminations and providing a voltage gate for the graphene sample. The barriers are uniform along the graphene edge, so the voltage and the resulting current distribution are also homogeneous. The entire structure is normally illuminated by a plane wave with magnetic field along the $y$-axis to excite the TM SPPs. For the present case, we assume two stacked graphene sheets to enhance the interaction between the incident plane wave and the 2D material. We approximate the effective conductivity of this compound structure as that of a single layer with a larger effective conductivity, $\sigma_{eff} = 2 \sigma$, as usually done in the literature \cite{N2Graphene,Hassani_PRA}. A longitudinal voltage generates drifting electrons along the $x$-axis.

As discussed in the previous section, the presence of drifting electrons in the graphene sheet changes its optical conductivity and breaks the symmetry between forward- and backward-propagating SPPs. While the previous section considered a graphene sheet on an idealized, infinite, grating, more realistic scenarios can be modeled and studied using the finite element method implemented, for example, in a commercial software, such as COMSOL Multiphysics \cite{COMSOL}. In the presence of drifting electrons, the graphene's conductivity becomes nonlocal, namely, it depends on the wavevector, and it should therefore account for the wavenumber of the space harmonics of the mode propagating on the considered periodically modulated structure. In particular, the space-dependent conductivity of a graphene sheet on a periodically modulated substrate can be calculated from the nonlocal $\boldsymbol{k}$-dependent conductivity as discussed in Ref. \cite{Current_controlled} :
\begin{equation}
	\sigma( x-x', \omega ) = \sum_n e^{ik_n(x-x')} \sigma (k_n, \omega),
\end{equation}
where $ k_n = 2\pi n /P + k_0\cos (\theta) $, $ P $ is the grating period, $ \theta  $ is the angle of incidence measured from the interface, and $ \sigma (k_n, \omega) = [\omega / (\omega - k_n  v_d)]  \sigma_0( \omega - k_n v_d )  $. In COMSOL, graphene can be modeled as a conducting boundary condition with surface current $\boldsymbol{j}(x, \omega) = \int_{x'} \sigma( x-x', \omega ) \boldsymbol{E}(x', \omega) dx'$, where $ \boldsymbol{E}(x', \omega) $ is the tangential component of the electric field along the surface and the integration is over the graphene sheet. By combining these two equations we get the following expression:
\begin{equation}
	\boldsymbol{j}(x, \omega) = \sum_n e^{ik_n x} \left \{ \int_{x'}  e^{-ik_n x'} \frac{\omega}{\omega - k_n  v_d}   \sigma( \omega - k_n v_d )  \boldsymbol{E}(x', \omega) dx' \right \},
\end{equation}
which can then be coded into COMSOL to model the induced current on the drift-biased nonlocal graphene sheet on a periodically modulated substrate. %This technique allows us to consider drifting electrons on graphene by defining a non-zero $v_d$. 
Our numerical investigations show that considering the $n= -20  \rightarrow +20$ space-harmonic terms is enough to guarantee convergence. If the structure is infinite, accurate results can be obtained by modeling one unit cell and using periodic boundary condition; however, this method is also approximately valid for finite but large enough structures, as in Fig. \ref{fig2}(a). In this situation the spatial integral in the above equation is over the entire graphene sheet, not just a unit cell. 

Using the described method, we analyzed the light absorption by graphene plasmons supported by the structure in Fig. \ref{fig2}(a), for different values of drift velocity. Since the considered structure has zero transmission and only specular reflection at the considered frequencies, the absorptance is calculated as $ 1 - \left | S_{11} \right |^2 $, where $S_{11}$ is the reflection coefficient. The results are shown in Fig. \ref{fig2}(b) wherein the absorption peaks appear when the normally incident light resonantly excites a momentum-matched space harmonic of the SPP supported by the graphene sheet. It is clear that in the presence of drifting electrons the SPP absorption peak splits into two distinct resonances related to SPPs propagating along and against the drifting electrons. The emergence of these asymmetric surface modes can be confirmed by looking at the dispersion diagram. Figure \ref{fig2}(c) shows the SPP dispersion curves for the same structure in Fig. \ref{fig2}(a) but extended to infinity. The inset shows a zoomed-in view for the third space-harmonic around the light line. Forward- and backward-propagating modes cross the $ k=0 $ axis (normal incidence) at two different frequencies, which correspond to the two absorption peaks in Fig. \ref{fig2}(b). The small discrepancy in these frequencies is due to the fact that in Fig. \ref{fig2}(c) the graphene sheet is assumed to be infinite with no terminations. 

The magnitude of the electric field along the graphene surface is shown in Figs. \ref{fig2}(d, e, f) at the absorption peak frequencies for the non-biased and current-biased cases. In the absence of drifting electrons the symmetric forward- and backward-propagating SPPs produce equal and symmetric field hot-spots at the two ends of the graphene sheet. However, when graphene is biased with a drift current, the asymmetry between forward- and backward-propagating modes leads to asymmetric hot-spots at the two ends. Figure \ref{fig2}(e) shows the lower-frequency absorption peak in Fig. \ref{fig2}(b), in the presence of drift current. At this frequency, we have a stronger backward-propagating mode (Fig. \ref{fig2}(c)), therefore the hot-spot is more intense on the left termination, and is also more than twice stronger than in the reciprocal case. In contrast, the higher-frequency absorption peak in Fig. \ref{fig2}(b) is related to the forward-propagating mode (see the inset of Fig. \ref{fig2}(c)). This leads to a stronger and more intense field hot-spot on the right end of the terminated graphene structure, as shown in Fig. \ref{fig2}(f). In the next section, we show that such asymmetric, intense hot-spots offer an opportunity to strengthen nonlinear interactions, as exemplified by the large enhancement of third-harmonic generation.

\section{Enhanced Nonlinear Effects}

The system shown in Fig. \ref{fig2}(a) can tightly confine the electric field along the graphene sheet, especially on the silicon corrugations and at the opaque barrier terminations. The field enhancement is even more pronounced at the opaque barriers when graphene is biased with drifting electrons, as shown in Figs. \ref{fig2}(e) and \ref{fig2}(f). As discussed in the Introduction, the combination of the large nonlinear susceptibility of graphene with the dramatic electric field enhancement in the current-biased nonreciprocal system considered here can lead to a particularly strong nonlinear response at THz frequencies.

To demonstrate the enhanced nonlinear optical effects stemming from the proposed drift-biased nonreciprocal graphene system, the third harmonic generation (THG) nonlinear process was studied. This is a common and well-studied third-order nonlinear effect that can transfer a part of the incident radiation power at the fundamental frequency $ \omega $ to its third-harmonic (TH) frequency $ \omega_{TH} = 3\omega $. This process is usually extremely weak but in our configuration it is substantially boosted due to the strong field enhancement combined with the inherently large nonlinear properties of graphene. The THG process will introduce an additional nonlinear term in the surface-current modeling formalism of graphene given by the formula $ J_{NL}(x,3\omega) = \sigma^{(3)} (x, \omega) E(x, \omega)^3 $, where $ \sigma^{(3)} $ is graphene’s third-order nonlinear surface conductivity and $E$ is the local electric field at the fundamental frequency \cite{Chatzidimitriou}. The third-order nonlinear surface conductivity of graphene is also dependent on the frequency and is given by the formula \cite{Cheng,Chris_9}: $ \sigma^{(3)} = \frac{i \sigma_0 (\hbar v_F e)^2}{ 48\pi (\hbar \omega)^4 } T(\frac{\hbar \omega}{2 \mu_c}) $, where $ \sigma_0 = e^2/4\hbar $, $ T(x) = 17G(x) - 64G(2x) + 45G(3x) $, $ G(x) = \mathrm{ln} |(1+x)/(1-x)| + i\pi H(|x|-1) $ and $ H(z) $ is the Heaviside step function. Similar to the linear graphene conductivity, $\sigma^{(3)}$ is also expected to become nonlocal due to the presence of drifting electrons and the resulting Doppler frequency shift \cite{Trull}. However, this nonlocal effect will be extremely weak because the amplitude of the nonlinear conductivity is already orders of magnitude lower than its linear counterpart $ (  \sigma^{(3)} \ll \sigma ) $ and, as a result, nonlinear nonlocal effects can be safely neglected in our computations. Furthermore, we also omit to include in our simulations the nonlinearity of silicon, since the THG stemming from a thin silicon layer is much weaker compared to graphene, as shown in the Supplemental Material. This is mainly due to the much higher third-order nonlinear susceptibility of graphene at THz frequencies compared to silicon \cite{Hendry}. 

The THG process strength is characterized by the conversion efficiency $ CE = P_{r,TH}/P_i $, where $ P_i $ is the incident power at the fundamental frequency and $ P_{r, TH} $ is the radiated TH power \cite{Nasari,Chirs_10}.  
To simplify the computational modeling, the input intensity is chosen to take relatively low values allowing only a negligible fraction of the incident power to be transferred to the TH frequency, an approximation usually referred to as the undepleted-pump approximation \cite{Meik,Scalora}. Additional details about the numerical approach used to simulate the THG process are provided as Supplemental Material.

\begin{figure}[h!]
	\begin{center}
		\noindent \includegraphics[width=6.5in]{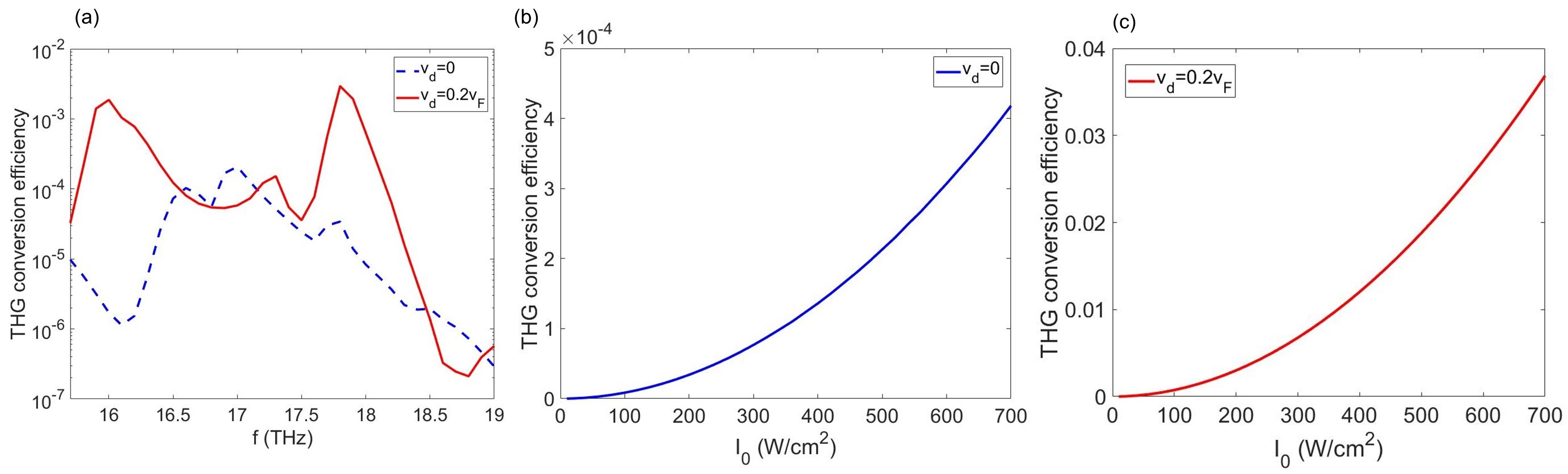}
	\end{center}
	\caption{(a) THG conversion efficiency for the reciprocal $ (v_d = 0 ) $ and nonreciprocal $ ( v_d = 0.2 v_F ) $ graphene structure in Fig. \ref{fig2}(a) as a function of the incident frequency $f$ when the input intensity $I_0$ is fixed to the low value of $200~ \mathrm{W/cm}^2$. (b)-(c) THG conversion efficiency as a function of the input intensity in the (b) reciprocal $ (v_d = 0) $ and (c) nonreciprocal $ (v_d = 0.2v_F) $ case plotted at a fixed frequency, $ f = 17.8  $ THz. }\label{fig3}
\end{figure}

The computed THG conversion efficiency for the system in Fig. \ref{fig2}(a) as a function of the incident frequency is shown in Fig. \ref{fig3}(a) with (red line) and without (blue line) a drift-current bias in the graphene sheet. With a drift velocity of $ v_d = 0.2 v_F $, the nonreciprocal THG conversion efficiency can be as high as $0.3 $ percents at the resonance frequencies presented before in Fig. \ref{fig2}(b) for the linear case. The used input intensity is extremely low in this case, approximately $200~ \mathrm{W/cm}^2$, which is a value easily attainable by current THz sources \cite{Tonouchi,Vitiello}. The nonreciprocal THG conversion efficiency for $ v_d = 0.2 v_F $ is approximately three orders of magnitude larger than in the reciprocal case with $ v_d = 0 $, when the results are compared around the two resonance peak frequencies. Finally, the THG conversion efficiency can be further improved by increasing the input intensity, as shown in Figs. \ref{fig3}(b) and \ref{fig3}(c) for the reciprocal and nonreciprocal cases, respectively, where the efficiency is computed at the nonreciprocal  resonance peak located at $f = 17.8 $ THz. Nonlinear damping leading to saturation and nonlinear absorption were not considered in our work, since the used input power level was very low. At such low power levels, saturation was not present in relevant harmonic generation experiments based on graphene structures \cite{graphene_fermions,grating_graphene}.

 %The rapid increase of the THG conversion efficiency originates from the nonlinear nature of the THG process, where the radiated TH power is tunable and proportional to the cubic power of the incident intensity at the fundamental frequency. Notable, the THG conversion efficiency in Fig. \ref{fig3}(a) is slightly higher in the intermediate frequency of $f=15 $ THz in the reciprocal case, however is approximately an order of magnitude lower compared to the THG conversion efficiency peaks at the two resonance frequencies in the nonreciprocal case.     

\section{Conclusion}

DC electric currents can be induced on graphene with ultra-high electron drift velocity, comparable to graphene's Fermi velocity. This provides an additional mechanism to control the propagation properties of surface plasmon-polaritons on graphene. In particular, the presence of drifting electrons makes graphene's optical conductivity nonlocal and nonreciprocal, a property that allows lifting the forward/backward degeneracy in SPP propagation along or against the electrons stream. Leveraging this property, we have theoretically and computationally showed that by applying a DC current on a terminated graphene structure, it is possible to generate controllable, asymmetric, and intense field hot-spots where the electric field is strongly localized and enhanced at the terminations. Since the third-order nonlinear polarization is proportional to the cubic power of the local electric field, we have used this strategy to boost graphene's nonlinear optical response and we have showed that, under realistic values of drift velocity, the THG conversion efficiency can be as high as 0.3 percents at the plasmon resonance frequencies. Our theoretical findings further demonstrate the potential of current-biased nonreciprocal systems to enable enhanced and anomalous light-matter interactions. We believe graphene plasmonics represents an ideal platform to realize these effects in practical systems.

\section*{Funding}

F.M. was funded by the Office of Naval Research (N00014-22-1-2486) and the Air Force Office of Scientific Research (FA9550-22-1-0204). C. A. is partially funded by the Office of Naval Research (N00014-19-1-2384) and National Science Foundation (2224456 and 2212050)

\begin{suppinfo}
	
The supplemental material provides more details about the numerical method used to simulate the THG process, the electric field distribution at the third-harmonic frequency, the effect of removing the opaque barriers in the considered graphene structure, and a short discussion on the nonlocal properties of graphene's conductivity. 
	
\end{suppinfo}

\end{document}